\documentclass[3p,times,procedia]{elsarticle}

\usepackage{ecrc}
\usepackage[utf8]{inputenc}
\usepackage[english]{babel}
\usepackage{siunitx}
\usepackage[super]{nth}
\usepackage[colorlinks]{hyperref}
\usepackage{longtable}
\usepackage{mathtools}
\usepackage{array}
\usepackage{setspace}
\usepackage{tabularx}
\usepackage{diagbox}
\usepackage{multirow}
\usepackage{graphicx}
\usepackage{amsmath}
\usepackage{array}
\usepackage{lscape} 
\usepackage[labelfont=bf]{caption}
\usepackage{subcaption}
\usepackage{afterpage}

\volume{00}

\firstpage{1}

\runauth{P. Borges et al.}

\jid{****}

\usepackage{amssymb}

\usepackage[figuresright]{rotating}
\usepackage{placeins}

\usepackage[utf8]{inputenc}
\usepackage[english]{babel}
\usepackage{siunitx}
\usepackage[super]{nth}
\usepackage{imakeidx}
\usepackage{blindtext}
\usepackage{booktabs}
\usepackage{graphicx}
\usepackage{multirow}
\usepackage{amsmath}
\usepackage{natbib} 
\setcitestyle{round,authoryear}
\makeindex[name=person,title={Index of persons}]
\makeindex
\usepackage{float,lscape}
\usepackage{pdflscape}
\usepackage{booktabs}

\usepackage{lineno}
\linenumbers

\makeindex
\usepackage{xr-hyper}

\begin{document}
\nolinenumbers
\begin{frontmatter}

\dochead{}

\title{Acquisition-invariant brain MRI segmentation with informative uncertainties}


\author{
Pedro Borges$^{*,1,2}$,
Richard Shaw$^{1,2}$,
Thomas Varsavsky$^{1,2}$,
Kerstin Klaser$^{2}$,
David Thomas$^{3}$,
Ivana Drobnjak$^{1}$,
Sebastien Ourselin$^{2}$,
M Jorge Cardoso$^{2}$}

\address{$^1$Department of Medical Physics and Biomedical Engineering, UCL, UK \\ $^2$School of Biomedical Engineering and Imaging Sciences, KCL, UK \\ $^3$Dementia Research Centre, UCL, UK \\ $^*$Corresponding author: p.borges.17@ucl.ac.uk
}

\begin{abstract}
Combining multi-site data can strengthen and uncover trends, but is a task that is marred by the influence of site-specific covariates that can bias the data and therefore any downstream analyses. Post-hoc multi-site correction methods exist but have strong assumptions that often do not hold in real-world scenarios. Algorithms should be designed in a way that can account for site-specific effects, such as those that arise from sequence parameter choices, and in instances where generalisation fails, should be able to identify such a failure by means of explicit uncertainty modelling. This body of work showcases such an algorithm, that can become robust to the physics of acquisition in the context of segmentation tasks, while simultaneously modelling uncertainty. We demonstrate that our method not only generalises to complete holdout datasets, preserving segmentation quality, but does so while also accounting for site-specific sequence choices, which also allows it to perform as a harmonisation tool.

\end{abstract}

\end{frontmatter}

\section{Introduction}

The largely non-quantitative nature of MRI means that it is significantly more susceptible to site effects compared to quantitative imaging modalities, as scanner or acquisition differences can result in a signal that can vary by orders of magnitude. Information is derived from the relative contrast between tissues rather than the value of the signal in said tissues. 

The substantial soft-tissue contrast of MRI makes it the tool of choice in a myriad of applications, especially in the field of neuroimaging. Different sequence and sequence parameters choices will result in the emphasis of different tissues, and these are deliberately chosen depending on the task at hand. There is therefore a high demand for algorithms that can process such varying data to a satisfactory degree without falling victim to the site-based biases embedded in the images themselves, given that data arising from different sites can have been acquired using non-standardised protocols.

Convolutional neural networks (CNNs) have achieved state of the art results in a multiplicity of medical imaging tasks including image-to-image translation, segmentation, and classification. CNNs, however, are susceptible to overfitting to the data regime in which they were trained and evaluated, resulting in poor generalising performance. This is due to the fact that the data used for training may not exhibit overlapping characteristics with the holdout set, the set of data that may have arisen, for example, from a different site that may have employed a different scanning protocol. This can be ameliorated with a robust augmentation scheme, but standard augmentation pipelines cannot replicate contrast differences between regions without further modelling, leading to inconsistent biomarker extraction~\citep{Shinohara}.

This leads us to a discussion on methods that attempt to tackle this contrast conundrum. Probabilistic generative models~\citep{AshburnerFriston2005} are widespread in their use, but are limited by their strict label intensity distributions and underlying assumptions. Multi-atlas fusion methods~\citep{Sabuncu2010} are likewise popular, but suffer from prolonged processing times owing to their registration-base nature.

CNNs have also been employed to address this problem. Billot et al.~\citep{Billot2020a} synthesise multi-contrast images using a Bayesian segmentation model that operates on label atlases, using these to train networks that should be able to adapt to images of any contrast. Pham et. al~\citep{10.1007/978-3-030-59520-3_1} designed a method whereby contrast robustness is attained by training a segmentation-synthesis model. A standard segmentation network is trained, which is then used to segment an image of unseen contrast. This (initially sub-optimal) segmentation is then used to train a synthesis network that is applied in combination with the observed contrast images' labels to regenerate these contrast images, which in turn serve to retrain the segmentation network, and the process is repeated. Crucially, there is no explicit modelling of the actual physics of acquisition and how it relates to contrast, and therefore no accounting for how these differences might result in segmentations that deviate from the actual anatomical ground truth.

It is important to also acknowledge works that approach the problem of multi-site harmonisation. For example, Combat is a Bayesian method that models site effects both additively and multiplicatively, allowing data harmonisation while preserving biological variability~\citep{Johnson2007}. Harmonisation has also been taken on with CycleGANs~\citep{Zhu2017, Zhao2019} and domain adaptation approaches~\citep{Dinsdale2020}. 

In our previous work~\citep{Borges2020}, we proposed a method that explicitly models the physics of the acquisition to design networks that can become invariant to the process by which they were acquired. This method made use of multiparametric MR maps (MPMs), which contain voxelwise quantitative MR parameter information combined with MR sequence simulations to generate images of various contrasts which are used to train networks privy to the sequence parameters used to generate the synthetic images. The work showed that networks trained in this fashion can achieve greater segmentation consistencies across a wide range of sequence parameters without incurring detriments to the segmentation quality. This work did not, however, offer out-of-distribution analyses, nor was their work validated extensively on real, external datasets.

Recently, uncertainty estimation has featured increasingly in deep learning works, as it grants us greater insight about both our data and the predictions provided by our networks. Acquisition parameters change tissue contrast and noise, to such an extent where certain choices of parameters can result in very uncertain segmentations, due to the lack of contrast. We can therefore leverage uncertainty estimation to measure the model's ability to perform a task as a function of the acquisition parameters. We choose to model both epistemic and heteroscedastic aleatoric uncertainty ~\citep{Kendall}. Epistemic uncertainty relates to the uncertainty in the model, while heteroscedastic aleatoric uncertainty relates to the uncertainty intrinsic to the data.

We significantly extend our previous work~\citep{Borges2020} and introduce several improvements that significantly contribute towards acquisition-invariance, namely: the translation of proposed simulation framework into a full dynamic data augmentation pipeline, the introduction of a contrast stratification loss, and the modelling of uncertainty. We hypothesise that these modifications should allow for a greater extrication of the physics of the acquisition and the underlying anatomy, and a greater ability to extrapolate to unknown regions of contrast and parameter space. This is evaluated by means of analysing segmentation consistency across in and out of distribution samples within subjects, as well as how the uncertainty-derived volumetric uncertainties differ between methods. Furthermore, we investigate how our method performs on a harmonisation task consisting of real multi-site data by analysing the consistency of predicted longitudinal tissue trends across sites.

\section{Methods}

In our previous work~\citep{Borges2020} we proposed a physics-aware segmentation training pipeline that took as input simulated MPM-based MR volumes in combination with the parameters used to generate said volumes to produce segmentations. These were shown to maintain greater volumetric consistencies in comparison to baseline methods when evaluated on images containing a wide range of tissue contrasts. We also put forward a "Physics Ground Truth" (PGS) label creation model to generate the tissue segmentations used to train these networks. We described the PGS as a true anatomical ground truth, as it is derived directly from the quantitative MPMs, and therefore not affected by the biases that a segmentation created from a qualitative image would have introduced. We continue using this PGS, by sourcing literature values for the tissues of interest, grey matter (GM), white matter (WM), and cerebrospinal fluid (CSF) and fitting a Gaussian mixture model for each on the quantitative values in the MPMs. All volumes are skull-stripped to ensure that no further extra-cranial tissue considerations are required. We propose improvements to our previous methodology, further validating our method on out of distribution samples aided by uncertainty derived errors, and evaluating our method on a multi-site harmonisation task to assess generalisability to a real, multi-contrast, dataset.

\subsection{Network architecture}

We adopt the nn-UNet architecture~\citep{Isensee2018} owing to its widespread adoption for segmentation tasks in literature. As in Borges et el.~\citep{Borges2020}, the acquisition parameters used to generate the images in any one batch are passed via two fully connected layers of length 40, before tiling the output of said layers both  following the second convolutional layer, and following the second to last convolutional layer. We posit that knowledge of the physics can enrich the features learned in the encoding part of the network instead of constraining the physics knowledge to the end layers of the network alone.

Networks are trained with a batch size of four, consisting of 3D patches, each of size $128^3$, uniformly sampled from the synthetically generated images. Networks are trained until convergence, which is met when seven epochs with no improvement in the validation metric have passed. We define the validation metric as the sum of dice score and volumetric coefficient of variation (calculated per tissue, across images in the batch), averaged across the three tissues. MONAI~\citep{MONAI} and TorchIO~\citep{Perez-Garcia2020} and PyTorch are used for all implementations.


\subsection{Stratification and batch homogeneity}

Intra-subject segmentation volume consistencies across multiple contrasts can be considered as a surrogate for acquisition parameter invariance. We propose some changes to the original methodology to further enforce this consistency. In particular, we leverage the fact that, for a single subject, the PGS segmentations are constant for every simulated realisation of the images (regardless of the choice of sequence and sequence parameters used to simulate images for that subject), as the underlying biology is unchanged. If a batch only contains realisations from a single subject, and the patch location is identical for all samples in the batch, then the labels for this batch are also identical. If using single subject batches, we can therefore add a constraint to the batch feature maps to enforce similarity between them. This comes in the form of a stratification ($L_2$) loss over all the feature maps in the penultimate layer of the network. We term this loss \textit{stratification} because its inclusion forces features to be the same, therefore stratifying style and content.

\subsection{Casting simulation as an augmentation layer}

The static equation simulation approach follows that described in Jog et al.~\citep{JOG}, specifically, we make use of the MPRAGE and SPGR equations described there. These approximate the signal, per voxel, given its intrinsic MR parameters ($T_1$, $T_2^{*}$, $PD$) and the chosen sequence parameters, which depend on the specific sequence being modelled. The absence of a temporal component is why they are termed static. The static signal for voxel $x$ for a MPRAGE sequence is:

\begin{equation} \label{eq:eqMPRAGE}
b_M(x) = G_SPD(x)\Bigg(1-\frac{2e^{\frac{-TI}{T_1(x)}}}{1+e^{\frac{-(TI+TD+\tau)}{T_1(x)}}}\Bigg),
\end{equation}
\newline
Accordingly, for a SPGR sequence:
\begin{equation} \label{eq:eqSPGR}
b_S(x) = G_SPD(x)sin{\theta}\frac{1-e^{-\frac{TR}{T_1(x)}}}{1-\cos{\theta}e^{-\frac{TR}{T_1(x)}}}e^{-\frac{TE}{T_2^{*}(x)}},
\end{equation}

For the sequence specific parameters, $G_S$ denotes the scanner gain, $TI$ is time between the inversion recovery pulse and
the first RF readout pulse, $TR$ the repetition time, $TE$ the echo time, $\tau$ the echo spacing time, and $TD$ the delay time. $G_S$ is a multiplicative factor that is assumed to be constant for all voxels, so is chosen to remain constant for all simulations. Note that the proposed static equation model is an approximation of the imaging process which ignores the local MRI dynamics, but allows for sufficiently-realistic and fast simulations necessary for CNN model training.

In our earlier work,~\citep{Borges2020} the synthetic image creation process is part of pre-processing; a set number are pre-generated according to a set parameter interval and are then used for training. We propose casting the static equation simulation process as an augmentation. In this fashion, the network takes in a protocol type, a range of relevant protocol parameters that are sampled from, and MPMs. Per iteration, a single MPM is selected, and by sampling from the range of protocol parameters N times, N simulated volumes are generated to make up the batch. This is in accordance with our aforementioned batch stratification modification, which therefore means that the selected patch for each of these samples resides in the exact same space.

\begin{figure}[htb]
\centering
\includegraphics[width=.9\textwidth]{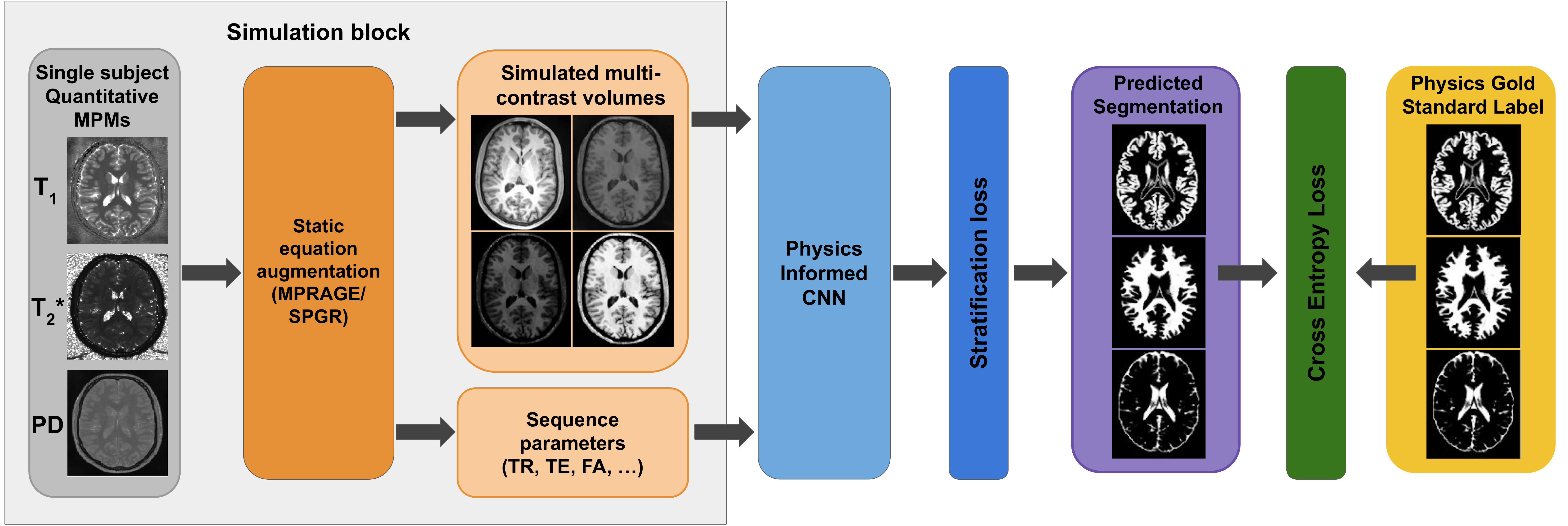} 
\caption{The training pipeline with proposed new additions of single subject batch stratification and accompanying $L_2$ feature maps loss, and training time image simulation.}
\label{fig:figPipeline}
\end{figure}

This eliminates the need for having to prepare the data in advance, and allows for a greater dynamic exploration of the physics parameter space. Fig.~\ref{fig:figPipeline} shows the training pipeline, featuring all aforementioned modifications.

In addition, We employ standard MR-related training time data augmentation, namely bias field and noise, to improve generalisability.

\subsection{Uncertainty modelling}

Data acquired with different parameters and devices results in differing levels of image contrast and noise, thus affecting a model's ability to segment a target region-of-interest; this effect can be characterised by the model's uncertainty. Modelling uncertainty allows us to obtain volumetric error bounds on our network outputs, which we can use to compare methods, as well as model performance with in and out of distribution samples. We choose test-time dropout with Monte Carlo sampling to model epistemic uncertainty. Via random network neuron activation we can sample from a series of subnets, each of which will produce different segmentation results, which approximates Bayesian posterior sampling~\citep{Gal2015}. The dropout rate is set to 0.5 in all layers to maximise the variance of those layers' outputs, except for the first layer, where we opt for a value of 0.05 to not penalise the learning of low-level image features too harshly~\citep{Eaton-Rosen2018}. 

Heteroscedastic aleatoric uncertainty is explicitly modelled by means of loss attenuation as described by Kendall et. al~\citep{Kendall}. By modelling the network outputs as a linear sum between its logits and voxelwise uncertainty predictions and propagating this to the cross-entropy loss we arrive at a new formulation:
\begin{align} \label{eq:eqHeteroLoss}
\hat{x}_{i,t} = f_i^W + \epsilon_t, \qquad \epsilon_t \sim \mathcal{N}(0, (\sigma_i^W)^2) \\
\mathcal{L}=\sum_{i}log\frac{1}{T}\sum_{t}c_{i}(-\hat{x}_{i,t,c} + log\sum_{c'}e^{\hat{x}_{i,t,c'}}) 
\end{align}

$\hat{x}_{i,t}$ is the result of summing $f_i^W$, the voxelwise task logits, with samples taken from a normal distribution with mean zero and standard deviation equal to the voxelwise network prediction of $\sigma_{i}^W$. This loss requires the internal component of the loss to be averaged over $T$ samples. In order to predict $\sigma_{i}^W$, we concatenate an additional branch to our architecture immediately following the final upsampling layer. It follows the same architectural structure as the original segmentation branch, with the addition of a softplus activation layer after the last convolutional layer to ensure that all predicted standard deviations are positive.

Under this new formulation, the network is encouraged to assign a high uncertainty to harder to segment regions, and likewise inclined to assign a low uncertainty to those regions it finds easier to classify. This choice of modelling allows us to obtain multiple segmentations from a single input, where differences are presumed to arise only due to the heteroscedastic uncertainty, from which volumetric bounds can be derived.

\section{Experiments and Results}

\subsection{Data and experimental details}
\subsubsection{Data}

Our data consists of quantitative multi-parametric volumes from 27 subjects, originating from a young onset Alzheimer disease dataset (YOAD)~\citep{Foulkes2016}~\citep{Slattery2019}. Each volume is 4-dimensional, 1mm isotropic, matrix size 181 x 217 x 181 x 4. The quantitative MR parameters span the fourth dimension. These parameters consist of $R_1$, the longitudinal magnetisation relaxation rate, $R_2^{*}$, the effective transverse magnetisation relaxation rate, proton density (PD), and magnetisation transfer (MT). MT does not feature in the static equation models we employ, so we do not make use of the MT maps. MPM creation details are described in~\citep{Helms2008}.

\subsubsection{Simulation sequence details}

We seek to compare how our additions compare with the original work, and so we therefore train our networks using simulated images bearing the same parameter ranges and sequences explored therein. For MPRAGE, this entails: TI = [600-1200] ms; for SPGR this entails: TR = [15-100] ms, TE = [4-10] ms, FA = [15-75] degrees. \textit{Physics Gold Standard} labels are generated using the same approach, with the same parameters chosen for the Gaussian mixture models of each tissue.

\subsubsection{Results presentation}
When comparing models and solutions, statistical significance is ascertained via signed-rank Wilcoxon tests, carried out independently on the different metrics. Values in \textbf{bold} denote the statistically best models. In instances where models may outperform baselines but are not statistically significantly different from each other, we bold both.

\subsection{Annealing study: Robustness and quality analysis}

We frame this comparative work as an annealing study, whereby we evaluate how model performance changes with each subsequent contribution, verifying their contributions to the model's efficacy. To this end we train a series of networks: \textit{Baseline}, a vanilla nn-UNet that takes as input pre-generated data (as in the original work), \textit{Phys-Base}, the physics-informed style network as proposed by the original work, also taking as input pre-generated data, \textit{Phys-Strat}, equivalent to \textit{Phys-Base} with the added stratification loss across the feature maps in the mini-batch, also trained with pre-generated data, and \textit{Phys-Strat-Aug}, equivalent to \textit{Phys-Strat} with the shift from using pre-simulated images to training with our proposed physics augmentation pipeline, taking as input therefore 4D MPMs.

As in the original work, 121 images are simulated per sequence for those models that are trained with pre-generated data. For MPRAGE, equally spaced TI intervals between [600-1200] are chosen for image creation, while for SPGR the parameters are sampled randomly from the space of TR = [15-100] ms, FA = [15-75] degrees, and TE = [4-10] ms. Performance is evaluated in terms of volume consistency (using coefficient of variation, CoV) and dice score.

We posit that a truly physics-informed network should be able to adequately extrapolate to images generated with sequence parameter values that lie outside the range seen during training. To this end, we additionally validate our models on such out of distribution (OoD) samples. For MPRAGE, this involves extending the TI range to [100-2000] ms, and for SPGR TR is expanded to [10-200] ms while TE is expanded to [2-20] ms, and FA is expanded to [5-90] degrees. The same performance metrics are employed to assess performance.

Dice scores and CoV for each experiment and tissue are shown in Table~\ref{tab:tabDice} and Table~\ref{tab:tabCoV}. CoV results showcase incremental improvements most evidently, with the largest single improvement arising from the addition of the stratification loss, which is congruent with expectations, as it is an explicit volume consistency related constraint.

\newcolumntype{d}{X}
\newcommand{\heading}[1]{\multicolumn{1}{c}{#1}}
\newcolumntype{P}[1]{>{\centering\arraybackslash}p{#1}}
\newcolumntype{Y}{>{\centering\arraybackslash}X}
\newcolumntype{v}{>{\hsize=.5\hsize}Y}
\newcolumntype{k}{>{\hsize=.75\hsize}Y}

\begin{table}[htb]
\centering
\begin{tabularx}{\linewidth}{|Y|v|v|v|v|v|v|v|v|}
\hline
\multirow{3}{*}{Experiments} & \multicolumn{8}{c|}{Sequence Dice Scores} \\ \cline{2-9}
& \multicolumn{4}{c|}{MPRAGE} & \multicolumn{4}{c|}{SPGR} \\ \cline{2-9}
& \multicolumn{2}{c|}{GM} & \multicolumn{2}{c|}{WM} & \multicolumn{2}{c|}{GM} & \multicolumn{2}{c|}{WM} \\ \cline{2-9}
& IoD & OoD & IoD & OoD & IoD & OoD & IoD & OoD \\
\hline

\multirow{2}{*}{\parbox{1.5cm}{Baseline}} &
0.966 & 0.956 & 0.953 & 0.934 &
0.878 & 0.872 & 0.893 & 0.873 \\
& (0.005) & (0.006) & (0.002) & (0.002) &
(0.021) & (0.008) & (0.023) & (0.011) \\

\hline

\multirow{2}{*}{\parbox{1.5cm}{Phys-Base}} &
\textbf{0.971} & 0.964 & \textbf{0.964} & \textbf{0.959} &
0.911 & 0.872 & 0.912 & 0.880 \\
& \textbf{(0.007)} & (0.009) & \textbf{(0.008)} & \textbf{(0.011)} &
(0.020) & (0.050) & (0.021) & (0.092) \\

\hline

\multirow{2}{*}{\parbox{1.55cm}{Phys-Strat}} &
\textbf{0.970} & \textbf{0.969} & 0.958 & \textbf{0.957} &
\textbf{0.929} & \textbf{0.911} & \textbf{0.922} & 0.894 \\
& \textbf{(0.005)} & \textbf{(0.005)} & (0.004) & \textbf{(0.005)} &
\textbf{(0.015)} & \textbf{(0.011)} & \textbf{(0.021)} & (0.040) \\

\hline

\multirow{2}{*}{\parbox{3cm}{Phys-Strat-Aug}} &
\textbf{0.971} & \textbf{0.971} &  \textbf{0.962} & \textbf{0.960} & 
\textbf{0.930} & \textbf{0.913} & \textbf{0.921} & \textbf{0.899} \\
& \textbf{(0.004)} & \textbf{(0.005)} & \textbf{(0.003)} & \textbf{(0.004)} & 
\textbf{(0.016)} & \textbf{(0.019)} & \textbf{(0.015)} & \textbf{(0.019)} \\
\hline
\end{tabularx}
\caption{Mean dice scores for \textit{Baseline}, \textit{Phys-Base}, \textit{Phys-Strat}, and \textit{Phys-Strat-Aug} on segmentation task, across inference subjects. All dice scores are calculated against a Physics Gold Standard. Standard deviations quoted in brackets. Bold values represent statistically best performances.}\label{tab:tabDice}
\end{table}

\begin{table}[htb]
\centering
\begin{tabularx}{\linewidth}{|Y|v|v|v|v|v|v|v|v|}
\hline
\multirow{3}{*}{Experiments} & \multicolumn{8}{c|}{Sequence CoVs (x$10^3$)} \\ \cline{2-9}
& \multicolumn{4}{c|}{MPRAGE} & \multicolumn{4}{c|}{SPGR} \\ \cline{2-9}
& \multicolumn{2}{c|}{GM} & \multicolumn{2}{c|}{WM} & \multicolumn{2}{c|}{GM} & \multicolumn{2}{c|}{WM} \\ \cline{2-9}
& IoD & OoD & IoD & OoD & IoD & OoD & IoD & OoD \\
\hline

\multirow{2}{*}{\parbox{1.5cm}{Baseline}}
& 6.39 & 22.50 & 14.94 & 51.12  &
61.91 & 170.10 & 32.57 & 158.93 \\

& (0.87) & (4.08) & (1.71) & (7.11) &
(7.61) & (31.32) & (11.98) & (16.83) \\

\hline

\multirow{2}{*}{\parbox{1.5cm}{Phys-Base}}
& 2.72 & 14.67 & 3.28 & 28.10  &
77.22 & 127.22 & 20.77 & 264.80 \\

& (2.12) & (7.30) & (2.01) & (3.98) &
(34.44) & (18.61) & (9.35) & (8.52) \\

\hline

\multirow{2}{*}{\parbox{1.55cm}{Phys-Strat}}
& 0.71 & 6.15 & 0.53 & \textbf{3.67} &
21.83 & 59.78 & 8.60 & 59.19 \\

& (0.23) & (1.51) & (0.25) & \textbf{(1.34)} &
(0.83) & (13.31) & (0.64) & (11.25) \\

\hline

\multirow{2}{*}{\parbox{3cm}{Phys-Strat-Aug}}
& \textbf{0.42} & \textbf{4.74} & \textbf{0.51} & \textbf{3.65}
& \textbf{15.76} & \textbf{28.88} & \textbf{7.12} & \textbf{44.78} \\

& \textbf{(0.22)} & \textbf{(1.30)} & \textbf{(0.23)} & \textbf{(0.62)} &
\textbf{(1.18)} & \textbf{(9.74)} & \textbf{(0.45)} & \textbf{(4.22)} \\

\hline

\end{tabularx}
\caption{Coefficients of variation (CoV) for \textit{Baseline}, \textit{Phys-Base}, \textit{Phys-Strat}, and \textit{Phys-Strat-Aug} on segmentation task, averaged across test subjects. Standard deviations quoted in brackets. Bold values represent statistically best performances.}\label{tab:tabCoV}

\end{table}

\begin{figure}[htb]
\centering
\includegraphics[width=1.0\textwidth]{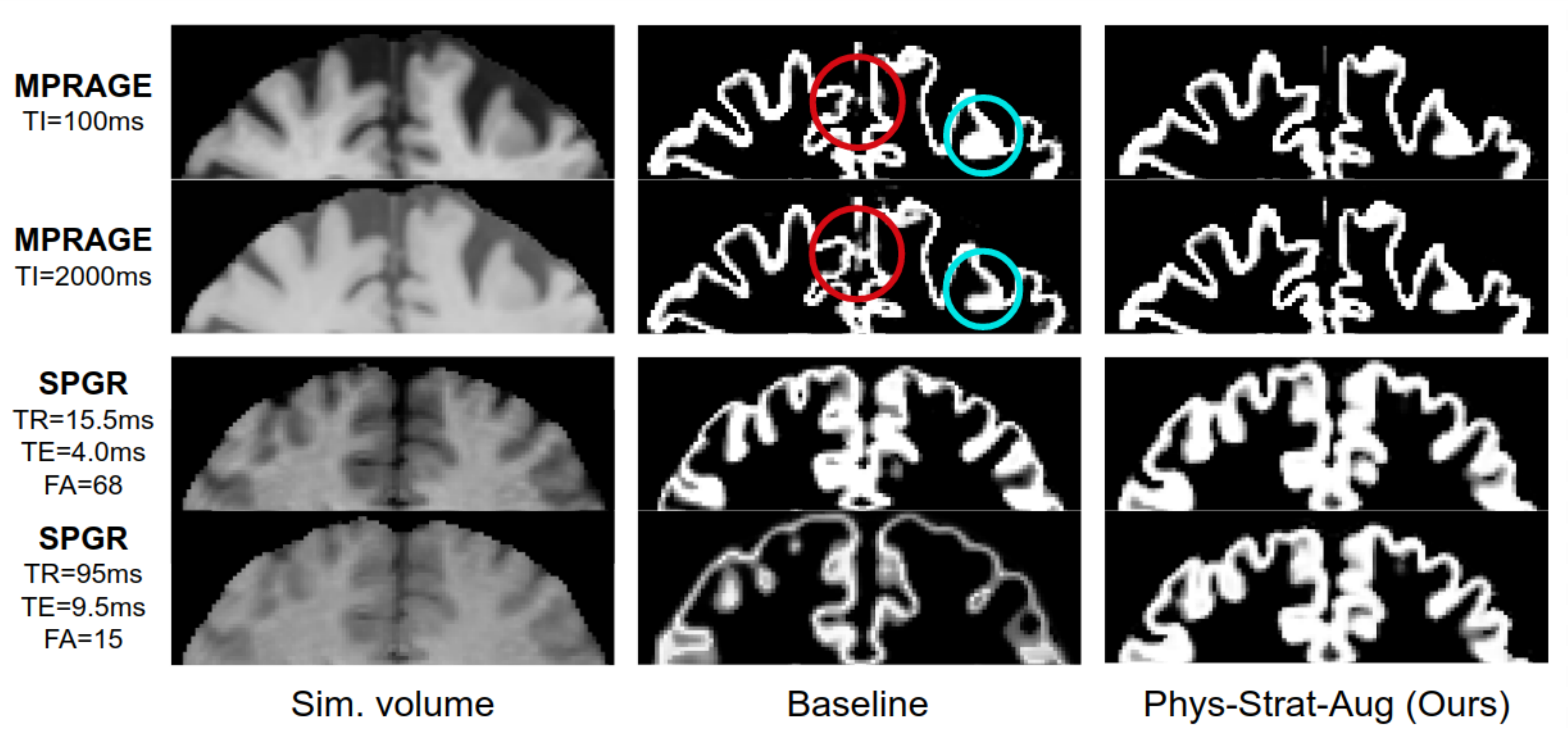} 
\caption{\textit{Baseline} and \textit{Phys-Strat-Aug} comparisons. Comparing out-of-distribution MPRAGE (Top two rows) and SPGR (Bottom two rows) GM segmentations from the proposed and baseline methods. Blue circles highlight examples of significant gyrus variability. Red circles denote regions of segmentation differences between protocols.} 
\label{fig:figIoD_OoD_Comps}
\end{figure}

We feature comparative qualitative segmentation results in Fig.~\ref{fig:figIoD_OoD_Comps} to showcase how consonant \textit{Phys-Strat-Aug}'s segmentations are across varying contrasts, compared to \textit{Baseline}. We circle specific regions where this is particularly evident. 

\subsection{Uncertainty measures and volumetric bounds}

We use the annealing study performances to inform network selection for our uncertainty investigation. \textit{Phys-Strat-Aug's} stands out as the best of the models, so we therefore train two sets of uncertainty-aware networks; an epistemic \textit{Phys-Strat-Aug} network, an epistemic \textit{Baseline} network, a heteroscedastic \textit{Phys-Strat-Aug} network, and a heteroscedastic \textit{Baseline} network. The intent is to ascertain how uncertainties deviate for a physics-informed network compared to a physics-agnostic counterpart, with respect to in and out of distribution samples.

To extract and calculate volumetric uncertainties we begin by sampling 50 segmentations from our epistemic networks, and 50 segmentations from our heteroscedastic networks, for in and out of distribution samples for MPRAGE and SPGR. Initial comparisons between networks made it apparent that the variance originating from heteroscedastic uncertainty was dwarfed by that originating from its epistemic counterpart, by several orders of magnitude. Eaton-Rosen et al.~\citep{Eaton-Rosen2018} also made this observation in their work that investigated the translation of uncertainty into quantitative error bounds, finding that the heteroscedastic contribution was negligible. As a result, our quantitative error analyses will henceforth focus solely on epistemic uncertainty.

\begin{figure}[htb]
\centering
\includegraphics[width=1.0\textwidth]{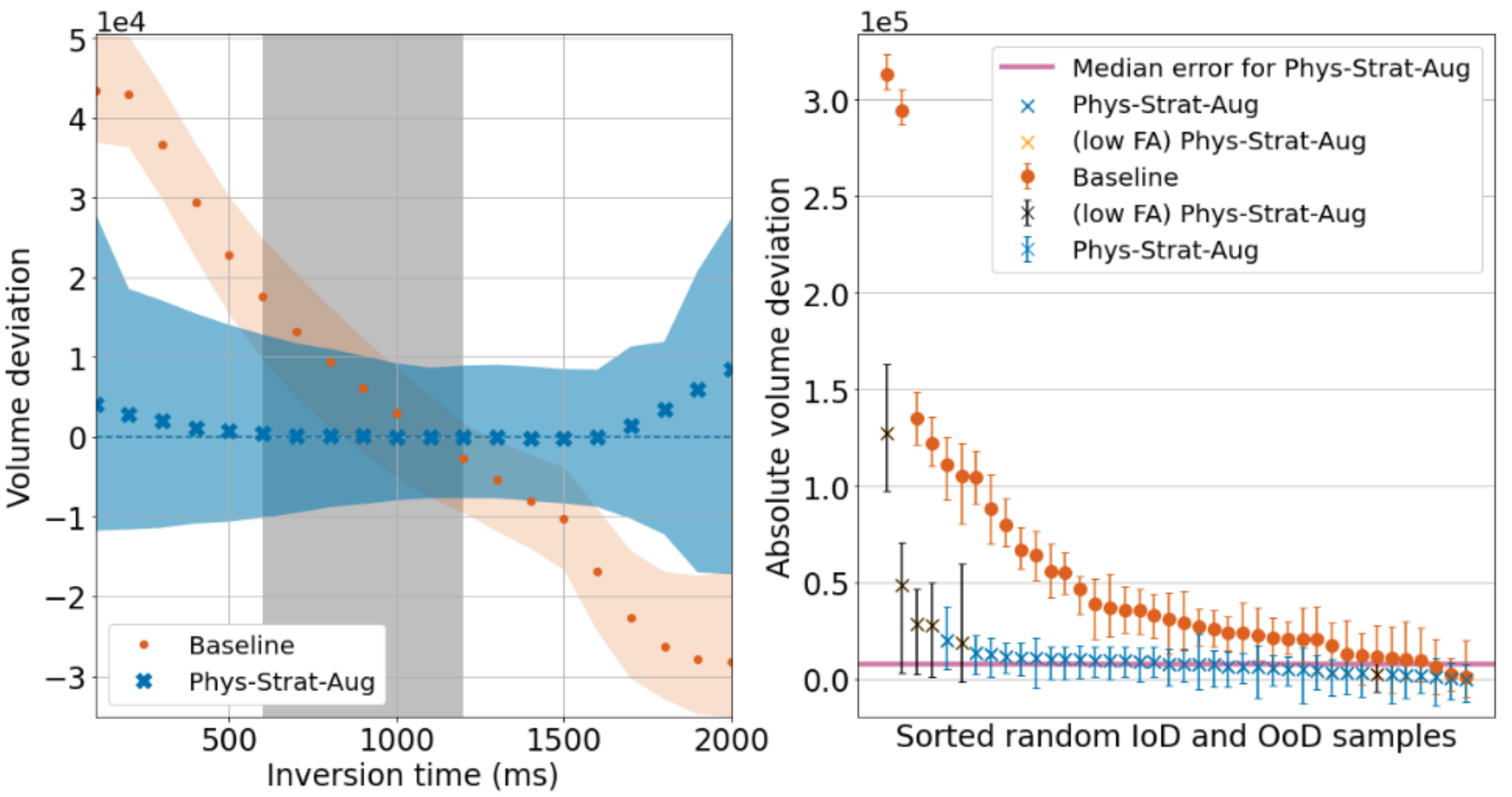} %
\caption{Comparing volume consistency for WM for \textit{Baseline} and \textit{Phys-Strat-Aug}, for example subject. Filled plots/ Error bars correspond to IQR volumes. Left: MPRAGE. The dashed grey region denotes the TI training time parameter range (600 - 1200 ms). Right: SPGR. Shown are the volume deviations from forty realisations of a single subject, where acquisition parameters were sourced at random from both the in an out of distribution parameter ranges. Points are sorted according to descending absolute volume deviation. The black points denote samples with FA lower than 10 degrees, for \textit{Phys-Strat-Aug}.} \label{fig:figMPRAGE_SPGR_Unc}
\end{figure}

To translate sampled segmentations into informative quantitative errors we follow the steps outlined in~\citep{Eaton-Rosen2018}. In summary, we first calculate, per tissue segmentation, per sample, voxelwise percentiles. We then calculate overall percentile volumes by aggregating voxels across Monte Carlo samples according to percentile value, before calibrating the cumulative distribution produced by these percentile volumes to ensure that confidence intervals reliably contain the expected ratio of ground truth values. Fig.~\ref{fig:figMPRAGE_SPGR_Unc} shows the volume variations for in and out of distribution inference samples, for white matter, for MPRAGE (Left) and SPGR (Right), accompanied by the interquartile range (IQR) volumetric error bounds. For SPGR, the parameter space from which samples are taken is three dimensional, so for visualisation purposes we order values according to the absolute volume error. We verify that in both instances \textit{Phys-Strat-Aug} exhibits significantly more consistent volumes, both in and out of distribution. 

Uncertainty-wise, we note that volumetric uncertainties are larger for out of distribution samples segmented by \textit{Phys-Strat-Aug} compared to \textit{Baseline}, for both MPRAGE and SPGR, where uncertainties remain more consistent throughout. This is especially evident when looking at the more extreme MPRAGE inversion time samples. 

By focusing on the outliers, we observe that for SPGR samples, uncertainty bounds are noticeably larger for \textit{Phys-Strat-Aug}. These outliers largely correspond to out of distribution samples, specifically images simulated with flip angles lower than ($10^{\circ}$). As with MPRAGE, the increased error for these samples in \textit{Phys-Strat-Aug} allows for most measurements to still overlap with the ground truth volume.

For SPGR, all the apparent outliers for \textit{Phys-Strat-Aug} have significantly larger associated errors, while this is not the case for the \textit{Baseline}. We observe that most outliers correspond to out of distribution samples using very low flip angles ($<10^{\circ}$, highlighted in black in the figure). Such images will be significantly less $T_1$-weighted, and therefore be less familiar to the models, in addition to having reduced contrast, resulting in poorer segmentation quality, so the observation that the physics-informed network's uncertainty around these samples is larger fits with expectations.

\subsection{Physics-driven multi-site harmonisation}

To assess the performance of the method in a multi-site research study setting, we use the ABIDE (Autism Brain Imaging Data Exchange) neuroimaging dataset~\citep{Martino2013}, which consists of structural and functional MR images from 19 sites, for 1112 subjects, 539 of which have been diagnosed with Autism Spectrum Disorder (ASD), and 572 of which are controls. We focus on the 11 sites that employed 3D MPRAGE acquisitions, resulting in 614 relevant images. Crucially, the sequence parameters employed across these sites differ, resulting in contrast differences between sites (in addition to other site-specific effects). The details of the acquisition parameters at each of these 11 sites can be found in Table~\ref{tab:tabABIDE}.

\afterpage{
\begin{landscape}
\begin{table}[b!]
\centering
\begin{tabularx}{\linewidth}{|k|k|k|k|k|k|k|k|k|k|}  
\hline
Site & Controls (m/f) & ASD (m/f) & Image acquisition & Voxel size ($mm^{3}$) & Flip angle (deg) & TR (ms) & TE (ms) & TI (ms) & BW (Hz/ Px) \\
\hline
\hline
CALTECH$^a$ & 15/4 & 15/4 & 3D MPRAGE & 1×1×1 & 10 & 1590 & 2.73 & 800 & 200 \\
\hline
CMU$^b$ & 10/3 & 11/3 & 3D MPRAGE & 1x1×1 & 8 & 1870 & 2.48 & 1100 & 170 \\
\hline
NYU$^c$ & 79/26 & 68/11 & 3D MPRAGE & 1.3×1×1.3 & 7 & 2530 & 3.25 & 1100 & 200 \\
\hline
OLIN$^d$ & 13/3 & 18/2 & 3D MPRAGE & 1×1×1 & 8 & 2500 & 2.74 & 900 & 190 \\
\hline
OHSU$^e$ & 15/0 & 15/0 & 3D MPRAGE & 1×1×1 & 10 & 2300 & 3.58 & 900 & 180 \\
\hline
UCLA$_1^f$ & 29/4 & 42/7 & 3D MPRAGE & 1×1×1.2 & 9 & 2300 & 2.84 & 853 & 240 \\
\hline
UCLA$_2^g$ & 12/2 & 13/0 & 3D MPRAGE & 1×1×1.2 & 9 & 2300 & 2.84 & 853 & 240 \\
\hline
PITT$^h$ & 23/4 & 26/4 & 3D MPRAGE & 1.1×1.1×1.1 & 7 & 2100 & 3.93 & 1000 & 130 \\
\hline
USM$^i$ & 43/0 & 58/0 & 3D MPRAGE & 1×1×1.2 & 9 & 2300 & 2.91 & 900 & 240 \\
\hline
YALE$^j$ & 20/8 & 20/8 & 3D MPRAGE & 1×1×1 & 9 & 1230 & 1.73 & 624 & 320 \\
\hline
\end{tabularx}
\caption{3D MPRAGE acquisition parameters for each relevant site in the ABIDE dataset~\citep{KucharskyHiess2015}. The scanner used at all sites was a Siemens Magnetom.}\label{tab:tabABIDE}
\caption*{$^a$ California Institute of Technology \\
$^b$ Carnegie Mellon University\\
$^c$ NYU Langone Medical Center, New York\\
$^d$ Olin, Institute of Living, Hartford Hospital\\
$^e$ Oregon Health and Science University\\
$^{f,g}$ University of California, Los Angeles\\
$^h$ University of Pittsburgh School of Medicine\\
$^i$ University of Utah School of Medicine\\
$^j$ Child Study Centre, Yale University}
\end{table}
\end{landscape}
}

By testing our physics-based segmentation methodology on this subset of the ABIDE dataset, our goals are twofold: To further demonstrate that networks trained on synthetic, MPM-based MR data, paired with a robust augmentation scheme, can generalise to data acquired at various different sites; and to show that the standardisation provided by accounting for the physics of acquisition at each site can result in harmonisation that is comparable to or exceeds the performance of existing harmonisation methods. 

The MPRAGE networks discussed thus far have been trained using images generated with only a varying TI. The MPRAGE static equation can also model the effects due to delay time, TD, and echo spacing time, $\tau$, which when combined with TI defines TR~\citep{wang2014optimizing}:
\begin{equation}
    TR = TI + TD + \tau
\end{equation}\label{eq:eqTR}
The MPRAGE static equation can therefore be re-written to incorporate TR directly:
\begin{equation} \label{eq:eqpTDMPRAGE}
b_M(x) = G_MPD(x)\Bigg(1-\frac{2e^{\frac{-TI}{T_1(x)}}}{1+e^{\frac{-TR}{T_1(x)}}}\Bigg),
\end{equation}
This allows us to directly incorporate the two main varying MPRAGE ABIDE site sequence parameters, TI and TR, into the augmentation simulation pipeline, which the network should learn to become robust to. Because TR is contingent on TI, we opt to model a pseudo parameter which we denote pTD, the sum of TD and $\tau$, which is added to TI to create TR.

To this end, we train three networks: A \textit{Phys-Strat-Aug} style network trained with MPRAGE images generated with TI = [600-1200] ms, and pTD = [500-1600] ms (which leads to TR = [1100-2800] ms), \textit{Phys-Aug-Base}, a network trained akin to \textit{Phys-Strat-Aug}, without the explicit passing of acquisition parameters during training, and \textit{CNN Baseline}, a baseline network trained with MPRAGE images generated with a single set of sequence parameters, to mimic the training of a standard CNN with images originating only from a single site. Additionally, we adopt the same robust augmentation scheme as outlined in Billot et al.~\citep{Billot2020a} to improve generalisability.

We used SPM12 (r7771, www.fil.ion.ucl.ac.uk/spm/), a widespread neuroimaging processing tool, running on MATLAB (R2019a, The MathWorks Inc., Natick, MA), to resample and rigidly co-register the images into a common space prior to any training. Additionally, we also employ SPM12 to generate grey matter, white matter, and CSF segmentations of these images, acting as our non-CNN baseline.

We compare harmonization performance with that offered by Combat~\citep{Johnson2007}. Combat models features as a combination of biological covariates (e.g.: Age, gender, pathology) and site effects, the latter of which can be subdivided into additive and multiplicative components. This allows Combat to regress out site effects while preserving biological variability. Combat has been shown to consistently outperform other harmonisation methods in various applications including multi-site cortical thickness measurement harmonisation~\citep{Fortin2018}, and multi-site diffusion tensor imaging data harmonisation~\citep{Fortin2017} and so we consider it to be the current gold standard. 

In the context of image segmentation, let $y_{ijv}$ denote the un-harmonized prediction for a specific tissue (grey matter, white matter, or CSF in our case) feature (voxels, volumes, etc.) $v$, for site $i$, and subject $j$. Combat models this value according to:
\begin{equation}
y_{ijv} = \alpha_v + X_{ij}\beta_v + \gamma_{iv} + \delta_{iv}\epsilon_{ijv}
\end{equation}\label{eq:eqCombat}
where $\alpha_v$ represents the average value of that feature across all sites and subjects, $X_ij$ represents the design matrix for biological covariates, $\beta_v$ the design matrix's corresponding featurewise coefficients, $\gamma_{iv}$ denotes featurewise additive site effects, $\delta_{iv}$ denotes the multiplicative featurewise site effects, and $\epsilon_{ijv}$ are the model residuals.

The harmonized feature values are derived via:
\begin{equation}
y_{ijv}^{Harm} = \frac{y_{ijv} - \alpha_v - X_{ij}\beta_v - \gamma_{iv}}{\delta_{iv}} + \alpha_v + X_{ij}\beta_v
\end{equation}\label{eq:eqCombatHarmonised}
While in certain applications the features employed can be individual voxels, this is not suitable for our use-case. The choice of feature would have to be congruous across all subjects, which would only be the case if all images were non-rigidly aligned to each other. Such an alignment process would prove destructive for the purpose of volumetric analysis however, making this unsuitable for our use-case. Furthermore, the computational cost to construct such a model for 1842 (614 subjects, three tissue segmentations each) 1mm isotropic 3D volumes would make this impractical. As such, we use the individual tissue volumes as our features, $v$ in our Combat models.

\begin{figure}[htb]
\centering
\includegraphics[width=1.\textwidth]{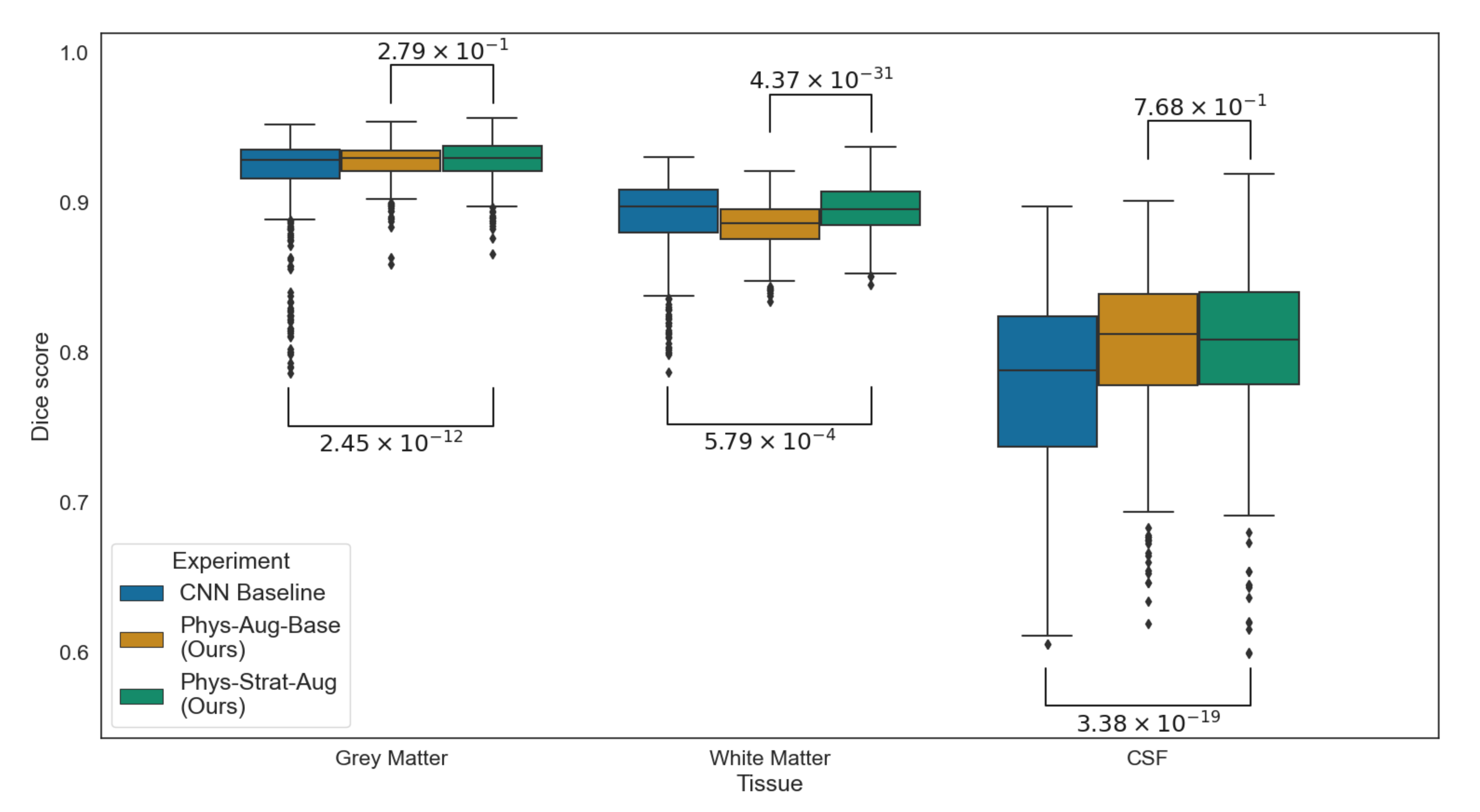} 
\caption{Dice scores of \textit{Phys-Strat-Aug}, \textit{Phys-Aug-Base}, and \textit{CNN Baseline} when compared against \textit{SPM}. Lines and attendant numbers denote p-values of scores between experiments, for each tissue.}
\label{fig:figDiceScores}
\end{figure}

As neural networks are often criticised by their unstable behaviour when applied to out of distribution data, to ascertain that the proposed method's segmentation is stable to OOD data, we calculate Dice scores between the segmentation outputs of the proposed models and SPM. SPM has seen widespread use for image segmentation due to its stability, combining mixture models, anatomical spatial priors, and MR-related intensity non-uniformity corrections to classify tissues, which justifies its use as a good non-CNN segmentation baseline. Fig.~\ref{fig:figDiceScores} shows the dice scores calculated between each of the neural networks' predictions and the corresponding SPM segmentation. We do not expect a perfect correspondence, especially when we expect any network exposed to our physics simulation-based methodology to account for the physics of acquisition, but these effects should not influence dice scores significantly. \textit{CNN Baseline} comparatively exhibits the greatest number of outliers, or failure cases, as well as a greater variance that skews towards lower dice scores. Dice scores for CSF are notably worse than for grey and white matter. 

\begin{figure}[htb]
\centering
\includegraphics[width=1.\textwidth]{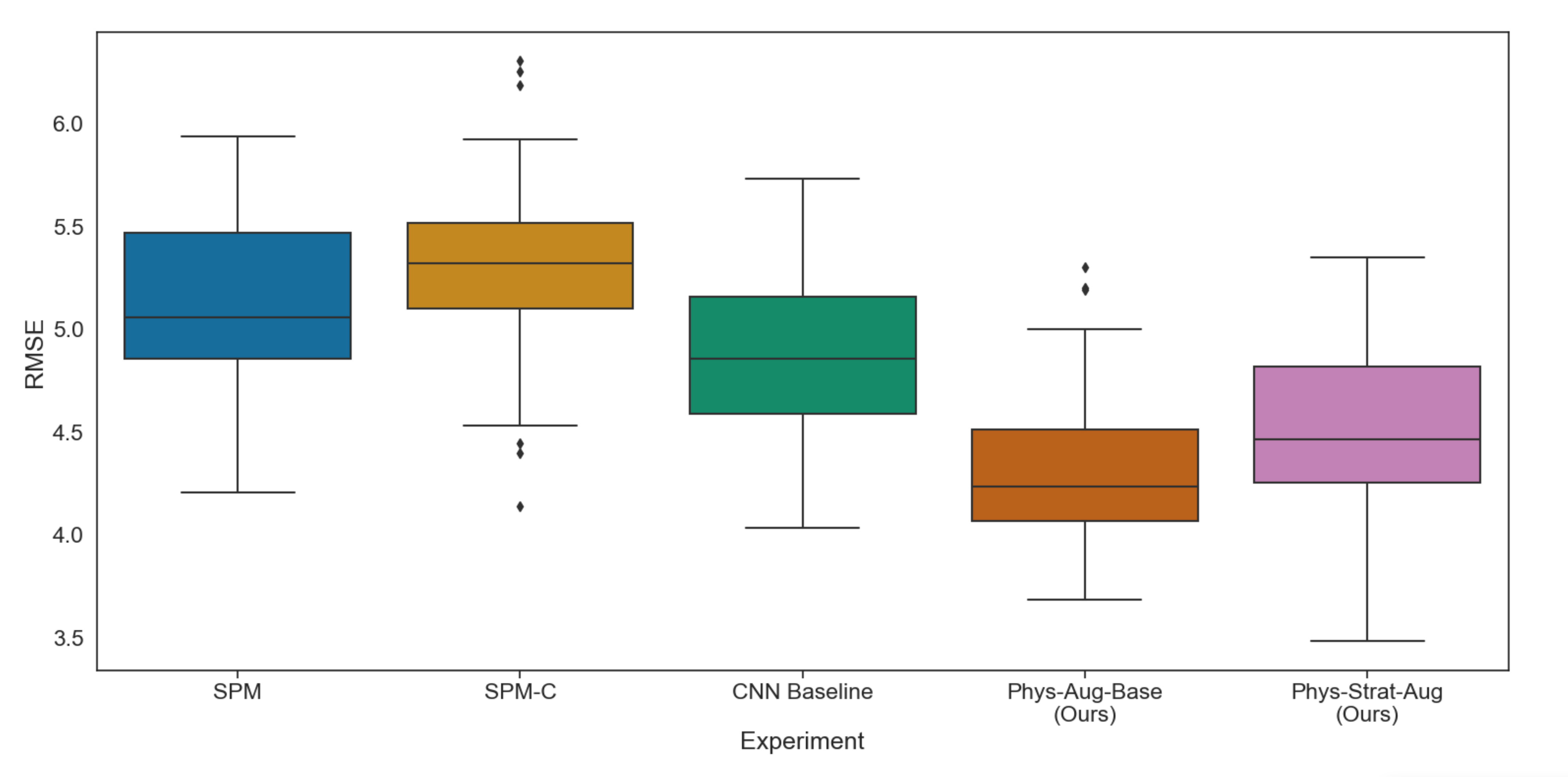}
\caption{Root-mean-square error boxplots for age prediction using linear regression, for each of the harmonisation experiments. Tissue volumes (GM, WM, and CSF) are used as model features.}
\label{fig:figAgeRMSE}
\end{figure}

We further investigate that age as a covariate has been preserved during the harmonisation process. This can be accomplished by fitting a linear age prediction model over the features for a subset of subjects, and testing this model on a holdout subset. The better the model's predictive ability the less the harmonisation process should've influenced the age-related biological variability. Fig.~\ref{fig:figAgeRMSE} showcases boxplots from the age regression models. Phys-Aug-Base boasts the lowest RMSE, while Combat-harmonised SPM (SPM-C) exhibits the worse performance.

The age distribution is not homogeneous across all sites, and so it is not reasonable to assume that age-based trends between age-heterogeneous sites distributions should be alike. We, therefore, partition the sites into two distinct groups based on a per-site mean age. The first we denote as "Young", where sites whose mean age is less than 16 are selected. The sites that fulfill this criteria are \textit{OHSU}, \textit{UCLA$_1$}, \textit{UCLA$_2$}, \textit{YALE}, and \textit{NYU}. The second partition we denote as "Old", where sites whose mean age is greater than 22 are selected. The sites that fulfill this criteria are \textit{CALTECH}, \textit{USM}, and \textit{CMU}. Under these criteria, \textit{PITT} is excluded.

Per model (for each site and tissue), per age-partitioned group, we use linear regression to fit a linear trend. Table~\ref{tab:tabHarmResults} shows the mean and standard deviations of these trends across sites, per group, tissue type, and experiment. We expect that, if the volumes are well-harmonised, the intercepts and gradients across all sites should be similar. The degree of similarity should be reflected in the standard deviation of these variables across all sites. 

\afterpage{
\begin{landscape}
\begin{table}[t!]
\centering
\begin{tabularx}{\linewidth}{|c|k|k|k|k|k|k|k|k|k|k|k|k|}
\hline
\multirow{3}{*}{Experiments} & \multicolumn{4}{c|}{Grey Matter} & \multicolumn{4}{c|}{White Matter} & \multicolumn{4}{c|}{CSF} \\ \cline{2-13} 
& \multicolumn{2}{c|}{"Young"} &
\multicolumn{2}{c|}{"Old"} & 
\multicolumn{2}{c|}{"Young"} &
\multicolumn{2}{c|}{"Old"} &
\multicolumn{2}{c|}{"Young"} &
\multicolumn{2}{c|}{"Old"} 
\\ \cline{2-13}
& b & m (x$10^{-3}$) & b & m (x$10^{-3}$) & b & m (x$10^{-3}$) & b & m (x$10^{-3}$) & b & m (x$10^{-3}$) & b & m (x$10^{-3}$) \\
\hline
\hline
\multirow{2}{*}{\parbox{2.25cm}{SPM (Baseline)}} & 0.6261 (0.0338) & -5.0864 (3.1716) & 0.5677 (0.0465) & -2.2552 (1.2661) & 0.2754 (0.0121) & 0.7202 (0.9218) & 0.2746 (0.0190) & 1.0360 (0.7280) & 0.0985 (0.0442) & 4.3661 (3.9279) & 0.1577 (0.0653) & 1.2192 (1.9581) \\
\hline
\multirow{2}{*}{\parbox{2.05cm}{CNN Baseline}} & 0.5584 (0.0382) & -5.9047 \textbf{(0.8932)} & 0.4435 (0.0773) & -2.0377 (1.2051) & 0.3848 (0.0236) & 0.3095 (0.7452) & 0.4481 \underline{\smash{(0.0972)}} & 0.9013 (0.3833) & 0.0569 \textbf{(0.0230)} & 5.5952 (1.3503) & 0.1084 (0.0340) & 1.1364 (1.3525) \\
\hline
\multirow{2}{*}{\parbox{2.225cm}{Phys-Aug-Base}} & 0.5353 \textbf{(0.0140)} & -3.7060 \textbf{(0.8859)} & 0.5000 (0.0174) & -1.8534 \textbf{(0.4316)} & 0.3736 (0.0208) & 1.0884 (0.7026) & 0.3923 (0.0227) & 0.6958 \textbf{(0.2525)} & 0.0911 \textbf{(0.0254)} & 2.6176 \textbf{(1.1613)} & 0.1077 (0.0308) & 1.1575 \textbf{(0.5582)} \\
\hline
\multirow{2}{*}{\parbox{2.25cm}{Phys-Strat-Aug}} & 0.5475 \textbf{(0.0114)} & -3.8187 \textbf{(0.7150)} & 0.4962 (0.0265) & -1.7860 \textbf{(0.3925)} & 0.4048 (0.0210) & 1.2285 (0.5688) & 0.4236 (0.0227) & 0.7272 \textbf{(0.2677)} & 0.0477 \textbf{(0.0179)} & 2.5902 \textbf{(0.5908)} & 0.0802 (0.0244) & 1.0587 \textbf{(0.5326)} \\
\hline
\multirow{2}{*}{\parbox{1.3cm}{SPM-C}} & 0.6286 (0.0471) & -4.5689 (3.9178) & 0.5691 (0.0340) & -2.1296 (1.3734) & 0.2797 (0.0107) & 0.7158 (0.8518) & 0.2716 (0.0195) & 1.0877 (0.7771) & 0.0947 (0.0541) & 3.5818 (4.4873) & 0.1554 (0.0463) & 1.1982 (1.8874) \\
\hline
\multirow{2}{*}{\parbox{2.45cm}{CNN Baseline-C}} & 0.5590 (0.0384) & -6.0404 \textbf{(1.0361)} & 0.4574 (0.0320) & -2.0423 (1.1976) & 0.3799 (0.0097) & 0.2363 (0.8568) & 0.4145 (0.0538) & 0.9936 (0.5009) & 0.0609 \textbf{(0.0114)} & 5.8040 (1.4680) & 0.1336 (0.0345) & 0.8382 (1.3743) \\
\hline
\multirow{2}{*}{\parbox{2.575cm}{Phys-Aug-Base-C}} & 0.5333 \textbf{(0.0132)} & -3.7094 \textbf{(0.8908)} & 0.5012 (0.0133) & -1.8832 \textbf{(0.4345)} & 0.3738 (0.0201) & 1.0993 (0.7404) & 0.3885 (0.0231) & 0.7313 \textbf{(0.2887)} & 0.0932 \textbf{(0.0248)} & 2.5794 (1.1293) & 0.1098 \textbf{(0.0294)} & 1.1668 \textbf{(0.5296)} \\
\hline
\multirow{2}{*}{\parbox{2.55cm}{Phys-Strat-Aug-C}} & 0.5525 \textbf{(0.0111)} & -3.8440 \textbf{(0.8408)} & 0.5006 (0.0105) & -1.7847 \textbf{(0.4032)} & 0.3967 (0.0138) & 1.2319 (0.5663) & 0.4212 (0.0162) & 0.7639 \textbf{(0.2978)} & 0.0506 \textbf{(0.0142)} & 2.6277 \textbf{(0.6260)} & 0.0783 \textbf{(0.0216)} & 1.0160 \textbf{(0.5235)} \\
\hline
\end{tabularx}
\caption{Trends statistics across sites, per age-partitioned group, tissue type, and experiment. Standard deviations quoted in brackets. Bold values denote improvement over SPM-C.}\label{tab:tabHarmResults}
\end{table}
\end{landscape}
}

We make use of Levene's test to ascertain whether the standard deviations between each experiment's trends and those of \textit{SPM-C} are significant. We verify that \textit{Phys-Strat-Aug} and \textit{Phys-Aug-Base} exhibit statistically significantly lower inter-site standard deviations than SPM-C for grey matter and CSF, for most instances of slope and intercept. CNN Baseline on the other hand does not perform better than SPM-C in most cases, and in fact shows noticeably worse intercept standard deviations, in particular for white matter in the "Old" cohort. Additionally, the mean trend gradients predicted from CNN Baseline's segmentations present the greatest departure from its counterparts when looking at the "Young" cohort. 

We further investigate the changes incurred by fitting Combat models to the segmentation volume values of each of the CNN-based experiments. We note that fitting a Combat model on all the CNN-based model outputs results in decreased intercept standard deviations, but slightly worse gradient standard deviations. This is unsurprising, as accounting for the additive effects of site (i.e.: any linear discrepancies, which relate to the linear intercept) is the easier of the tasks to perform, but it seems to come at the cost of gradient variation. Unsurprisingly, CNN Baseline benefits the most from this process, as its site volumes displayed the greatest disparity in most cases. 

\begin{figure}[htb]
\centering
\includegraphics[width=1.\textwidth]{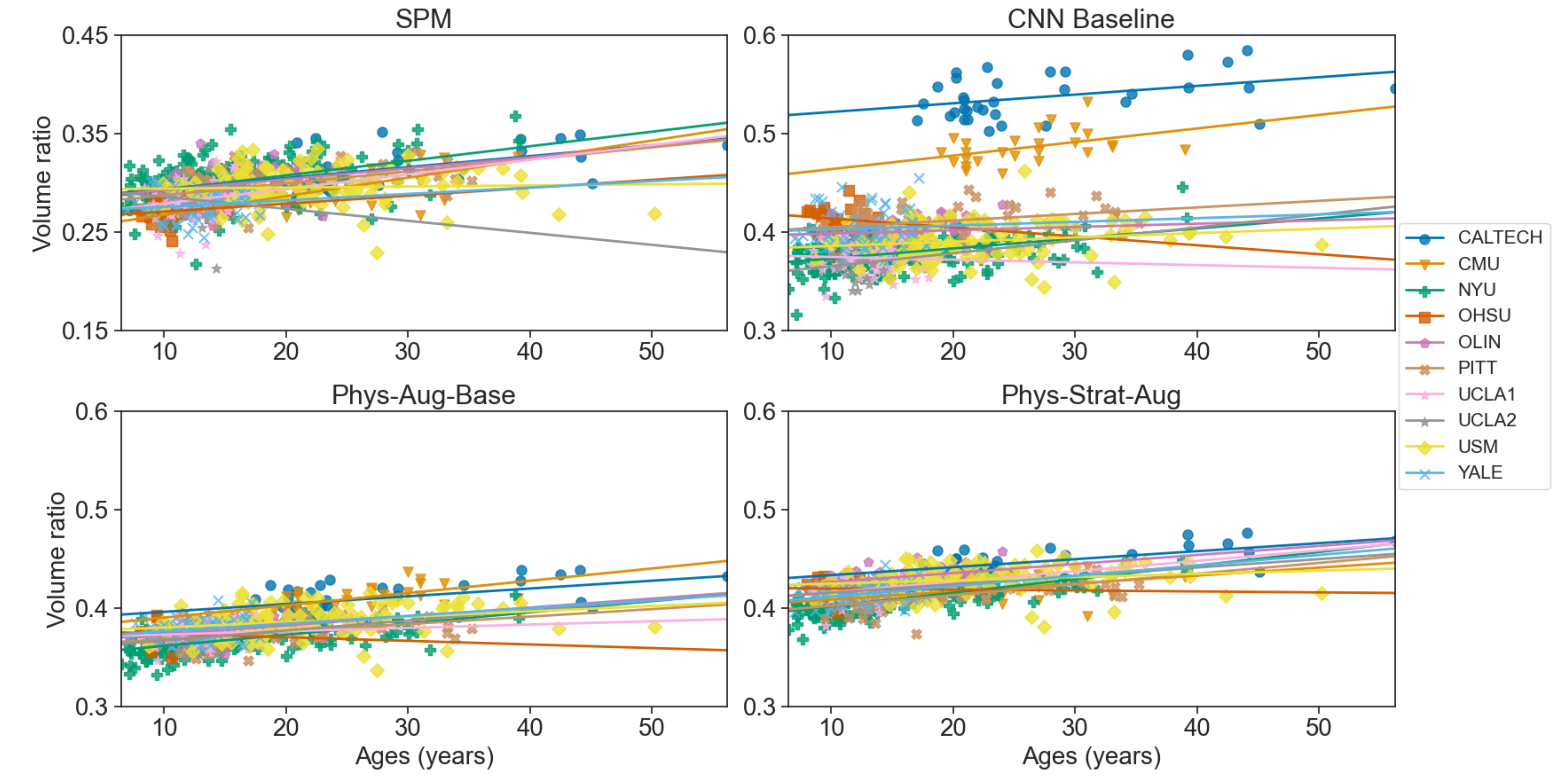}
\caption{White matter volume ratios for all MPRAGE subjects in the ABIDE dataset, for each of the experiments. Linear trends of best fit are calculated and shown, per site, extrapolated to the full dataset age range. Top left: SPM, Top right: \textit{CNN Baseline}, Bottom left: \textit{Phys-Aug-Base}, Bottom right: \textit{Phys-Strat-Aug}}
\label{fig:figCNN_WM_Plots}
\end{figure}

Fig.~\ref{fig:figCNN_WM_Plots} shows a scatter plot against age of white matter volume ratios for all subjects, for all the CNN-based experiments. Qualitatively, this figure clearly shows that \textit{CNN Baseline} has completely failed to generalise to two of the sites, as evidenced by their distinctly higher white matter volume ratios compared to other sites. This is not observed for \textit{Phys-Strat-Aug} or \textit{Phys-Aug-Base}, indicating that the physics augmentation pipeline has allowed these networks to aptly generalise to sites exhibiting different tissue contrasts. This is also reflected in Fig.~\ref{fig:figDiceScores}, which illustrates CNN Baseline's greater number of negative outliers, as well as its statistically significantly poorer performance compared to \textit{Phys-Strat-Aug}, for all tissues.

\subsection{Uncertainty-informed sequence optimisation}

In most neuroimaging research studies, MRI images are obtained as a mechanism to extract surrogate biomarkers of interest. These images are thus optimised to maximise signal and contrast (and subsequently good measurements) in the region of interest. As uncertainty is not only proportional to the degree of unfamiliarity but also to the difficulty of the task, an image which results in the lowest segmentation uncertainty of the region-of-interest should be optimal from the point of view of the surrogate biomarker of interest. 

We can therefore leverage uncertainty predictions to ascertain parameter choices or regions in parameter space that minimise said uncertainty. In theory images produced with these parameter pairs will be easiest for networks to segment, and should produce the most confident predictions. Furthermore, we can verify if those regions of minimal uncertainty overlap with typical sequence parameters sourced from real life studies.

To this end, we run inference on two uncertainty-aware networks, for each sequence of interest, SPGR and MPRAGE, on simulated volumes spanning the range of sequence parameters to be investigated. We then calculate the calibrated volumetric uncertainty for each of these volumes as outlined in previous sections, and aggregate the results in a contour plot. The range of parameters spanned for SPGR are (TR) = [5-100] ms, and flip angle (FA) = [5-90] degrees. Because we only deal with $T_1$-weighted images, whose contrast is largely dependant on TR, TE is ideally minimised. As such, it is fixed at 4 ms, Furthermore, its inclusion would increase the processing time multiplicatively. Akin to the MPRAGE networks trained for the ABIDE harmonisation, we vary both pTD and TI. The range of parameters spanned for MPRAGE are (pTD) = [200-2000], (TI) = [400-2000].

\begin{figure}[H]
\centering
\includegraphics[width=1.\textwidth]{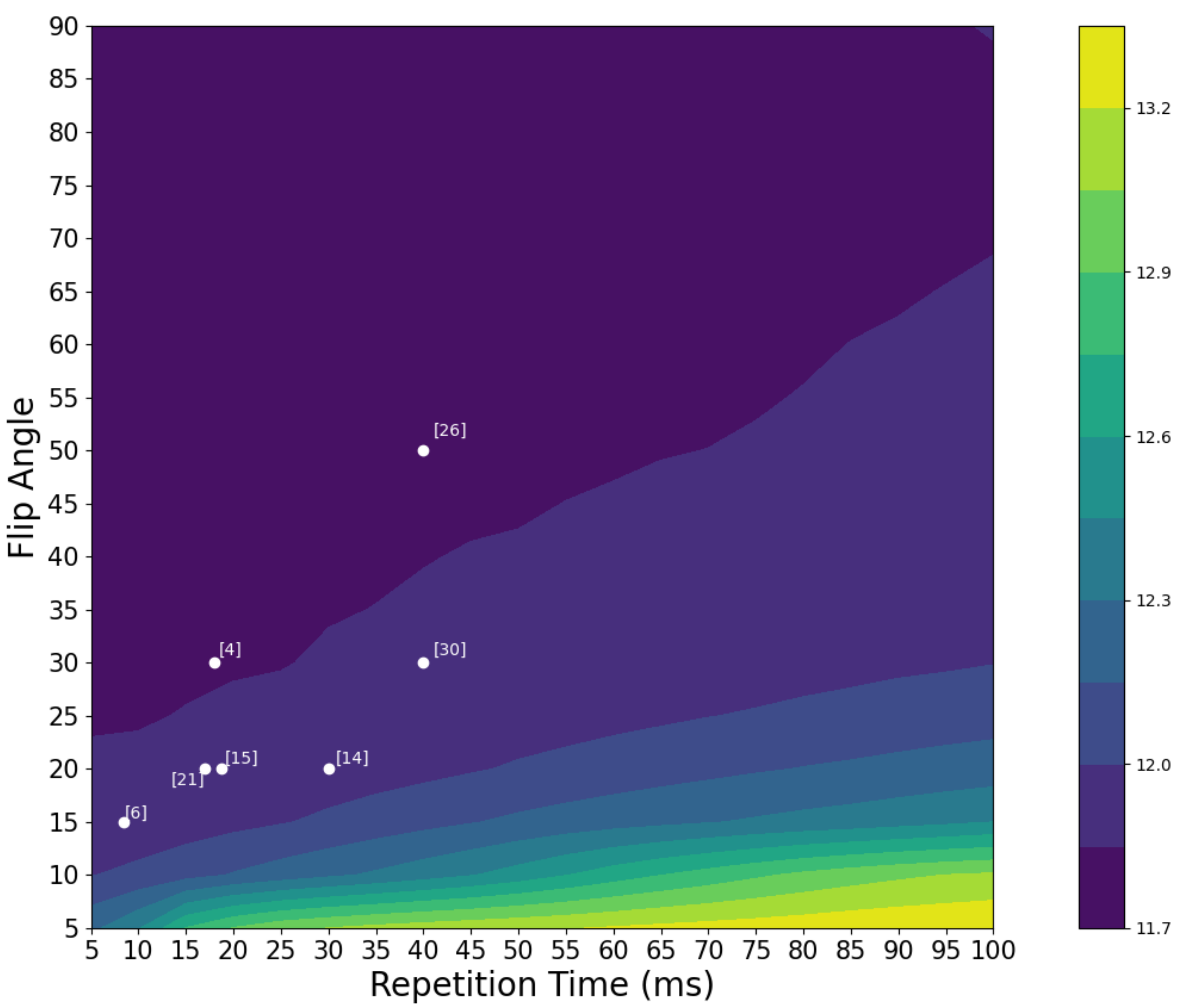}
\caption{Logarithmic SPGR uncertainty contour, averaged over all tissues and subjects. Scattered white points denote parameter choices sourced from relevant neuroimaging literature. Numbers enclosed in square brackets adjacent to each point denote the relevant reference: [4]:~\cite{Cocosco}, [6]:~\cite{DiMartino2014}, [14]:~\cite{Helms2008}, [15]:~\cite{Helms2009}, [21]:~\cite{Leow2007}, [26]:~\cite{Runge1991}, [30]:~\cite{Taki2011}}
\label{fig:figSPGR_Contour}.
\end{figure}

\begin{figure}[H]
\centering
\includegraphics[width=1.\textwidth]{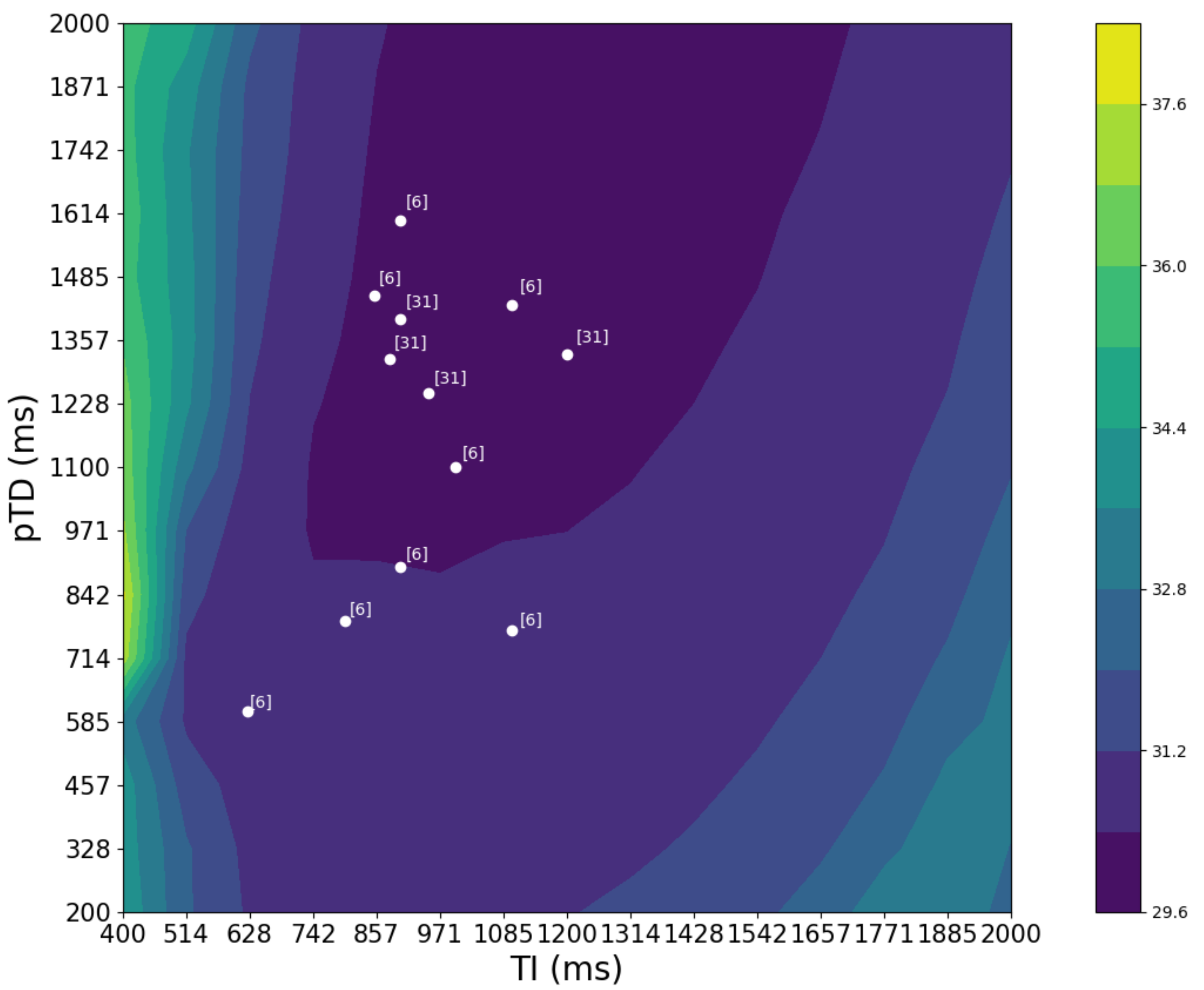}
\caption{Logarithmic MPRAGE uncertainty contour, averaged over all tissues and subjects. Scattered points denote parameter choices sourced from relevant neuroimaging literature. Numbers enclosed in square brackets adjacent to each point denote the relevant reference: [6]:~\cite{DiMartino2014}, [31]:~\cite{Wang2014}.}
\label{fig:figMPRAGE_Contour}
\end{figure}

Fig.~\ref{fig:figSPGR_Contour} and Fig.~\ref{fig:figMPRAGE_Contour} show the logarithmic volumetric uncertainty contours for SPGR and MPRAGE, respectively, averaged across all tissues and inference subjects. Though not shown, there is a significant overlap between subjects, which lends credence to the notion that the uncertainty is driven by the contrast between tissues, and should therefore be independent of subject-specific anatomy. The scattered points denote the sequence parameters used by various studies and trials, where the annotations correspond to relevant references to these. For MPRAGE there is a significant overlap between literature sequence parameters and regions of lowest uncertainty.

For SPGR this is also verified, though there is an absence in literature of $T_1$-weighted SPGR sequences employing larger ($>$ 50$^{\circ}$) flip angles and larger ($>$ 60 ms) repetition times. Higher repetition times are correlated with decreased $T_1$-weighting, and increased proton density weighting~\citep{Gras2013}. Additionally, if a good contrast can be attained at a lower $TR$ then it stands to reason that it would be selected since this would also save time by reducing the total acquisition time (which is proportional to $TR$).

The absence of higher flip angles in $T_1$-weighted SPGR studies can be explained in the context of Ernst angles, the flip angle at which the signal for a particular tissue is maximised given a certain TR and the $T_1$ properties of the tissue of interest. For the $TRs$ typically employed in $T_1$-weighted SPGR sequences, the Ernst angle for GM and WM lies in the range of [10-25]$^{\circ}$~\citep{Runge}. Maximising a single tissue's signal won't guarantee that the contrast between tissues is maximised, which can result in higher FAs being selected. Far from the Ernst angle for these tissues, the signal falls exponentially however, even if the relative contrast remains significant, which in turns results in a lower signal-to-noise ratio. Our simulation model does not explicitly account for this as a noise model is absent, but is well suited for inclusion as future work.

\section{Discussion and Conclusions}

In this work we demonstrated that with some well justified modifications to the training pipeline, a physics-informed network can achieve extremely constrained tissue segmentations across a wide range of contrasts, across all tissue types and investigated sequences, thus strengthening its harmonisation capabilities. 

Barring MPRAGE OoD WM results, \textit{Phys-Strat-Aug} boasts the statistically best performance. It benefits from the volume consistency enforcement of the stratification loss, as well as from the richer exploration of sequence parameter space, which should not only grant it greater invariance to the choice of acquisition parameters, but also greater generalisability owing to the functionally infinite training data at its disposal.

Furthermore, we also showed that it can suitably generalise to unseen domains, while maintaining volume consistency without compromising segmentation quality, and is validated by accurately quantifying the volumetric uncertainty. The uncertainty estimates further suggest that the incorporating knowledge of the acquisition scheme grants the model an additional level of safety, as volumetric uncertainties proved to be larger for out of distribution parameter generated images.

From our uncertainty-based sequence validation we verify that, for MPRAGE, there is a significant overlap between the sequence parameter region of lowest uncertainties assigned by our physics models and those sequence parameters employed in the literature for neuroimaging studies. This is a twofold boon, as it not only lends credence to the model by showcasing this congruence, but also could also allow for exploratory work involving newly developed sequences, which given their static equation analogue, can be used in combination with our uncertainty-aware networks to ascertain in advance those parameter combinations that might be most favourable for analyses. The SPGR counterpart showcased that while the minimum uncertainty overlap is verified, there are clearly unaccounted for variables that need to be considered, such as noise modelling. 

Our experiments on the multi-site ABIDE dataset showcase how our pipeline can provide harmonised segmentations for a multi-site volume study that out-compete baselines. Furthermore, we demonstrate how a naively trained, single-contrast baseline fails to generalise to multiple sites, providing less congruent trends, worse tissue volume ratio matching, and worse segmentation quality. When combined with the observation that the performance between \textit{Phys-Strat-Aug} and \textit{Phys-Aug-Base} is comparable, there exists the implication that the most relevant component in the pipeline is the multi-contrast physics augmentation block, though small further improvements do seem to be achieved by explicit incorporation of the sequence acquisition parameters.

We further emphasise how our physics-informed networks were able to generalise well to a complete holdout dataset composed of images originating from multiple different sites, acquired using different sequence parameters. This is in spite of the low number (27) of unique phenotypes involved in training and validating our models. We attribute this both to our physics augmentation pipeline, and to the robust augmentation scheme employed. However, as mentioned previously, the CSF Dice scores are significantly lower than their GM and WM counterparts. This can in part be attributed to how SPM's CSF maps include non-CSF tissue, such as dura mater, which are absent from the CNN-based experiments' CSF segmentations.

The method is admittedly limited to those sequences that can be aptly represented as a static equations, but we argue that at the very least, for the purposes of contrast agnosticism, a wide enough range of realistic contrasts can be generated with currently implemented sequences, which should allow for our method to generalise further. Future work will therefore involve investigating the exploration of means to aptly model MR artifacts such as noise, movement and $B_0$/$B_1$ inhomogeneities, to enhance our model's utility. 

\subsection{Credit Author Statement}

Pedro Borges is the primary author, having conceived and carried out the analyses, as well as having written the manuscript.

Richard Shaw aided in uncertainty-related discussions and implementations.

Thomas Varsavsky contributed towards the making of figures and revising the manuscript.

Kerstin Klaser contributed towards the making of figures and revising the manuscript.

David Thomas contributed to discussions involving the physics of MR acquisition, helped grant access to the YOAD, and revised the manuscript.

Ivana Drobnak contributed to discussions pertaining to MR simulation, and revised the manuscript.

Sebastien Ourselin contributed to funding.

M Jorge Cardoso contributed to the conception, development, and discussion of the analyses conducted in the work, and revised the manuscript.

\subsection{Data and code availability statement}

The YOAD was acquired at the National Hospital for Neurology and Neurosurgery, Queen Square. This dataset is not openly available.

Further details regarding the ABIDE data employed by this work, and the means to gain access to it, can be sought via \url{http://fcon_1000.projects.nitrc.org/indi/abide/}. Gaining access requires users to register with the NITRC and 1000 Functional Connectomes.

Code supporting the findings of this work, as well as a means to run some of our physics-informed models can be found in the following GitHub repository (\url{https://github.com/pedrob37/Phys_Seg}) under a permissive OpenSource license. 

\subsubsection{Acknowledgements}
This project was funded by the Wellcome Flagship Programme (WT213038/Z/18/Z) and Wellcome EPSRC CME \newline(WT203148/Z/16/Z). D.L.T. was supported by the UCL Leonard Wolfson Experimental Neurology Centre (PR/ylr/18575), the UCLH NIHR Biomedical Research Centre, and the Wellcome Trust (Centre award 539208).

The authors declared no potential conflicts of interest with respect to the research, authorship, and/or publication of this article.

\setcitestyle{numbers} 
\bibliographystyle{plainnat}
\bibliography{samplepaper}
\end{document}